\documentclass[sigconf]{acmart}

\AtBeginDocument{%
  \providecommand\BibTeX{{%
    \normalfont B\kern-0.5em{\scshape i\kern-0.25em b}\kern-0.8em\TeX}}}

\copyrightyear{2022}
\acmYear{2022}
\setcopyright{acmlicensed}\acmConference[ICSE '22]{44th International Conference on Software Engineering}{May 21--29, 2022}{Pittsburgh, PA, USA}
\acmBooktitle{44th International Conference on Software Engineering (ICSE '22), May 21--29, 2022, Pittsburgh, PA, USA}
\acmPrice{15.00}
\acmDOI{10.1145/3510003.3510062}
\acmISBN{978-1-4503-9221-1/22/05}

\usepackage{algorithm}
\usepackage{algorithmic}

\begin{document}
\renewcommand{\algorithmicrequire}{\textbf{Input:}} 
\renewcommand{\algorithmicensure}{\textbf{Output:}}
\newcommand{\etal}{{\textit{et al.}}}
\newcommand{\eg}{{\textit{e.g.}}}
\newcommand{\ie}{{\textit{i.e.}}}

\newcommand{\tabincell}[2]{\begin{tabular}{@{}#1@{}}#2\end{tabular}} 

\title{Bridging Pre-trained Models and Downstream Tasks for Source Code Understanding}



\author{Deze Wang}
\affiliation{%
	\institution{National University of Defense Technology, China}
	\country{}}
\email{wangdeze14@nudt.edu.cn}

\author{Zhouyang Jia}
\authornote{Zhouyang Jia and Shanshan Li are the corresponding authors.}
\affiliation{%
	\institution{National University of Defense Technology, China}
	\country{}}
\email{jiazhouyang@nudt.edu.cn}

\author{Shanshan Li}
\authornotemark[1]
\affiliation{%
	\institution{National University of Defense Technology, China}
	\country{}}
\email{shanshanli@nudt.edu.cn}

\author{Yue Yu}
\affiliation{%
	\institution{National University of Defense Technology, China}
	\country{}}
\email{yuyue@nudt.edu.cn}

\author{Yun Xiong}
\affiliation{%
	\institution{Fudan University}
	\city{Shanghai}
	\country{China}}
\email{yunx@fudan.edu.cn}

\author{Wei Dong}
\affiliation{%
	\institution{National University of Defense Technology, China}
	\country{}}
\email{wdong@nudt.edu.cn}

\author{Xiangke Liao}
\affiliation{%
	\institution{National University of Defense Technology, China}
	\country{}}
\email{xkliao@nudt.edu.cn}


\begin{abstract}
With the great success of pre-trained models, the pretrain-then-finetune paradigm has been widely adopted on downstream tasks for source code understanding. However, compared to costly training a large-scale model from scratch, how to effectively adapt pre-trained models to a new task has not been fully explored. In this paper, we propose an approach to bridge pre-trained models and code-related tasks. We exploit semantic-preserving transformation to enrich downstream data diversity, and help pre-trained models learn semantic features invariant to these semantically equivalent transformations. Further, we introduce curriculum learning to organize the transformed data in an easy-to-hard manner to fine-tune existing pre-trained models.

We apply our approach to a range of pre-trained models, and they significantly outperform the state-of-the-art models on tasks for source code understanding, such as algorithm classification, code clone detection, and code search. Our experiments even show that without heavy pre-training on code data, natural language pre-trained model RoBERTa fine-tuned with our lightweight approach could outperform or rival existing code pre-trained models fine-tuned on the above tasks, such as CodeBERT and GraphCodeBERT. This finding suggests that there is still much room for improvement in code pre-trained models.
\end{abstract}




\maketitle

\section{Introduction}
Large-scale models, such as BERT~\cite{Devlin2019BERTPO}, RoBERTa~\cite{Liu2019RoBERTaAR}, GPT-3~\cite{Brown2020LanguageMA}, T5~\cite{Raffel2020ExploringTL}, and BART~\cite{Lewis2020BARTDS}, have greatly contributed to the development of the field of natural language processing~(NLP), and gradually form the pretrain-then-finetune paradigm. The basic idea of this paradigm is to first pre-train a model on large general-purpose datasets by self-supervised tasks, \eg, masking tokens in training data and asking the model to guess the masked tokens. The trained model is then fine-tuned on smaller and more specialized datasets, each designed to support a specific task. The success of pre-trained models in the natural language domain has also spawned a series of pre-trained models for programming language understanding and generation, including CodeBERT~\cite{Feng2020CodeBERTAP}, GraphCodeBERT~\cite{Guo2021GraphCodeBERTPC}, PLBART~\cite{Ahmad2021UnifiedPF}, and the usage of T5 to support code-related tasks~\cite{Mastropaolo2021StudyingTU}, improving the performance of a variety of source code understanding and generation tasks.

However, pre-training a large-scale model from scratch is costly. Additionally, along with an increasing number of pre-trained models, how to effectively adapt these models for a new task is not fully exploited. In this paper, we try to take the first step to bridge large pre-trained models and code-related downstream tasks. Moreover, despite the success of existing pre-trained models for code-related tasks, these models have two potential issues. First, these models graft NLP pre-training techniques to understand the semantics of source code, however, the semantics of programming language and natural language are essentially different, and semantically equivalent source code may be in various syntactic forms. The second issue is that pre-trained models typically have at least millions of parameters, so when a pre-trained model is applied to downstream tasks with specialized datasets, there is a risk of overfitting because the model is over-parameterized for the target dataset. Many studies have also found that when the test set is different from the actual scene or the test set is slightly perturbed, various models for source code would make mistakes~\cite{Quiring2019MisleadingAA,Ramakrishnan2020SemanticRO,Yefet2020AdversarialEF}.

\begin{figure*}[t]                         
	\centering                              
	\includegraphics[width=\linewidth]{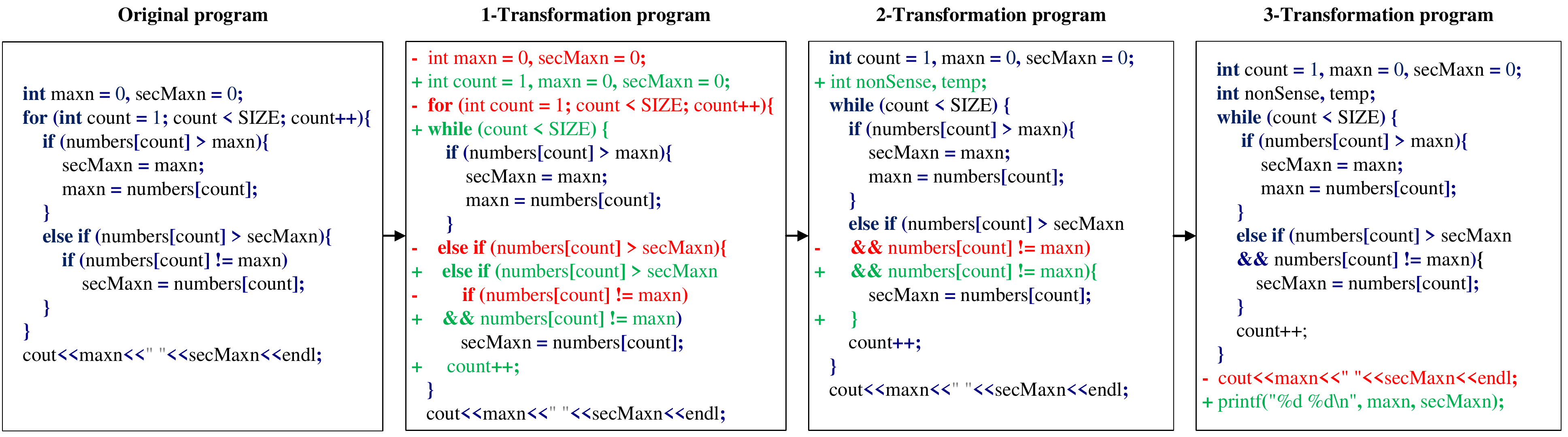} 
	\caption{An example of code transformation. $k-Transformation$ program represents the result of the original program after $k$ transformations. All four programs implement the function to find the maximum and second largest values in the array.}
	\label{example1}                              
\end{figure*}

To address the above issues, we design a lightweight approach on top of the existing pre-trained language model fine-tuning paradigm, that satisfies (1) extracting code semantic knowledge embedded in diverse syntactic forms and complementing it to pre-trained models, (2) reducing overfitting to the target dataset and being more robust in testing. In order to incorporate semantic knowledge of the programming languages into models, we employ data augmentation, which is mainly used to enrich the training dataset and make it as diverse as possible. There are many successful applications of data augmentation in the field of image processing, such as random cropping~\cite{Krizhevsky2012ImageNetCW}, flipping~\cite{Simonyan2015VeryDC} and dropout~\cite{Srivastava2014DropoutAS}. For code data, this paper considers semantic-preserving transformation. An example of code transformation is shown in Fig.~\ref{example1}, where the same program is transformed three times successively, keeping the semantics unchanged. Since the semantics of the original program are preserved, it is logical that the model should have the same behavior as the original program for the program generated by the transformation techniques. Moreover, it is cheap to leverage a source-to-source compiler~\cite{Aho2006CompilersPT} to perform semantic-preserving transformations on source code.

In this paper, we build our approach on a series of large-scale pre-trained models, including natural language pre-trained model RoBERTa and code pre-trained models CodeBERT and GraphCodeBERT, to bridge pre-trained models with downstream tasks for source code. We first construct semantic-preserving transformation sequences and apply them to original training samples, as in Fig.~\ref{example1}, to generate new training data and introduce code semantic knowledge into models. The transformation sequences make code transformations more complicated and could guide models to better learn the underlying semantics of the code. These training data are then fed to pre-trained models to fine-tune the models. Finally, we augment the test sets with the same augmentation techniques as the training sets to obtain multiple transformed test sets. To further reduce overfitting from the training process, we average the model performance on these test sets. Since our method averages the predictions from various transformation versions for any code snippet in test sets, the final predictions are robust to any transformation copy.

The transformed data significantly increase the data diversity, however, they can also be considered as adversarial examples compared to the original data~\cite{Ramakrishnan2020SemanticRO,Rabin2020OnTG}. Fig.~\ref{example1} shows the original program and programs after multiple code transformations. As the number of transformations increases, new tokens and syntactic forms are introduced, and the distribution of transformed data becomes more distinct from that of original data, making it more difficult to learn. To solve this issue, we introduce Curriculum Learning (CL)~\cite{Matiisen2020TeacherStudentCL} and present training examples in an easy-to-hard manner, instead of a completely random order during training. Many studies have shown that it benefits the learning process not only for humans but also for machines~\cite{Elman1993LearningAD,Krueger2009FlexibleSH}. The key challenge of CL is how to define easy and hard samples, and in this paper we propose two hypotheses and experimentally verify them to determine the learning order.

In our experiments, based on pre-trained models CodeBERT and GraphCodeBERT, our method significantly surpasses the state-of-the-art performance on algorithm classification, code clone detection and code search tasks. In the algorithm classification task, our approach improves 10.24\% Mean Average Percision~(MAP) compared to the state-of-the-art performance, and in the code clone detection task, using only 10\% of the randomly sampled training data, code pre-trained model CodeBERT fine-tuned with our approach outperforms the state-of-the-art model GraphCodeBERT normally fine-tuned with all training data. In the code search task, our method improves the state-of-the-art performance to 0.720 Mean Reciprocal Rank~(MRR). More impressively, to test whether our approach introduces additional semantic knowledge of source code for the model, we apply our approach to natural language pre-trained model RoBERTa and find that it even outperforms CodeBERT with 3.88\%
MAP on algorithm classification task and RoBERTa pre-trained with code on code search task, and has the same performance as CodeBERT on code clone detection task. The data, pre-trained models and implementation of our approach are publicly available at the link: \url{https://github.com/wangdeze18/DACL}.

The main contributions of our paper are as follows:
\begin{itemize}
	\item We design a lightweight approach on top of the existing pre-trained language model fine-tuning paradigm, to bridge pre-trained models and downstream tasks for source code. To the best of our knowledge, it is the first work in this direction.
	\item We apply our method to pre-trained models CodeBERT and GraphCodeBERT, and the augmented models dramatically outperform the state-of-the-art performance on algorithm classification, code clone detection and code search tasks.
	\item Our study reveals that for code-related tasks, without the need for heavy pre-training on code data, natural language models (\eg~RoBERTa) easily outperform the same models pre-trained with code, as well as the state-of-the-art code pre-trained models (\eg~CodeBERT) with the help of our approach.
\end{itemize}

\section{Preliminaries and Hypotheses}\label{sec2}

\subsection{Data Augmentation}

Data Augmentation (DA) is a technique to create new training data from existing training data artificially. Transformations were randomly applied to increase the diversity of the training set. Data augmentation is often performed with image data, where copies of images in the training set are created with some image transformation techniques performed, such as zooms, flips, shifts, and more. In fact, data augmentation can also be applied to natural language and code data. In this paper, our purpose of introducing data augmentation is to learn code semantics from semantic-preserving transformation, more specifically, to assist models in extracting and learning features in a way that are invariant to semantically equivalent declarations, APIs, control structures and so on.

In this paper, we exploit data augmentation not only for the training set but also for the test set. The application of data augmentation to the test set is called Test-Time Augmentation~(TTA)~\cite{Simonyan2015VeryDC,Moshkov2020TesttimeAF}. Specifically, it creates multiple augmented copies of each sample in the test set, has the model make a prediction for each, and then returns an ensemble of those predictions. The number of copies of the given data for which a model must make a prediction is often small. In our experiment, we randomly sample three samples for each piece of data from their augmented copies, take the average results as the result of the augmented perspective and add the results on the original dataset as the final results.

\subsection{Curriculum Learning}

The learning process of humans and animals generally follows the order of easy to difficult, and CL draws on this idea. Bengio \etal ~\cite{Bengio2009CurriculumL} propose CL for the first time imitating the process of human learning, and advocate that the model should start learning from easy samples and gradually expand to complex samples. In recent years, CL strategies have been widely used in various scenarios such as computer vision and natural language processing. It has shown powerful benefits in improving the generalization ability and accelerating convergence of various models~\cite{Guo2018CurriculumNetWS,Jiang2014EasySF,Platanios2019CompetencebasedCL,Tay2019SimpleAE}. At the same time, it is also easy-to-use, since it is a flexible plug-and-play submodule independent of original training algorithms.

There are two key points of CL, one is the scoring function and the other is the pacing function. The scoring function makes it possible to sort the training examples by difficulty, and present to the network the easier samples first. The pacing function determines the pace by which data is presented to the model. The main challenge is how to obtain an effective scoring function without additional labelling of the data.

\subsection{Hypotheses}\label{hypotheses}
We formulate two hypotheses about the scoring functions to determine the order of learning and conduct experiments to verify them. 

Many studies have shown that deep models for source code are vulnerable to adversarial examples~\cite{Quiring2019MisleadingAA,Ramakrishnan2020SemanticRO,Yefet2020AdversarialEF}. Slight perturbations to the input programs could cause the model to make false predictions. Therefore, it is natural for us to formulate the first hypothesis that \textbf{the augmented data are more challenging to learn than the original data for general models}.  We design an experiment to verify this hypothesis directly, as shown in Algorithm~\ref{algorithm1}. It shows the pseudocode to verify the impact of code transformation by comparing the performance of the model on a range of training set variants. The training set variants are generated by iterating the transformation functions on the original training set. (line 4-7) After the model is trained on the whole training set including all training set variants, we evaluate the model on different training set variants.

\begin{figure}[htbp]
	\centerline{\includegraphics[width=0.9\linewidth]{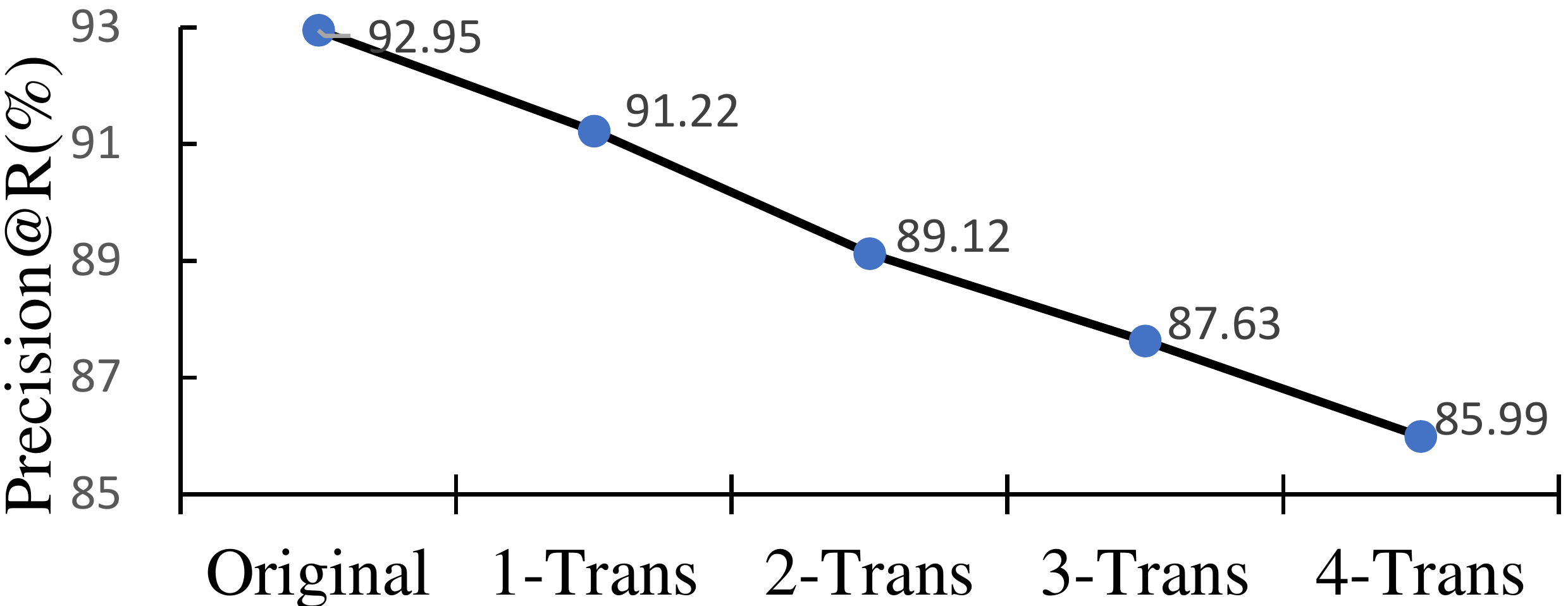}}
	\caption{The performance of CodeBERT on both original and augmented training sets for POJ104 dataset.}
	\label{fig3}
\end{figure}

We apply Algorithm~\ref{algorithm1} to the state-of-the-art model CodeBERT with benchmark dataset POJ104~\cite{Mou2016ConvolutionalNN}~(will be explained in \ref{datasett}). Fig.~\ref{fig3} shows the performance of CodeBERT for these training set variants. The model performs best on the original training set. The performance gets progressively worse as the number of transformations on the original dataset increases, which verifies that data augmentation would increase the difficulty of the training set and experimentally supports our hypothesis.

\begin{algorithm}[htp]
	\caption{Validation Algorithm for Hypothesis 1}
	\label{algorithm1}
	\begin{algorithmic}[1] 
		\REQUIRE Training set $D$, transformation functions $T_1,...,T_k$, model $M$
		\STATE {$\Gamma \gets \{D\}$, a set of training set variants}
		\STATE {$\Omega \gets \{\}$, a set of experimental results}
		\STATE {$D' = D$}
		\FOR {transformation $t \gets 1~...~k$}
		\STATE {$D' = T_t(D')$}
		\STATE {$\Gamma \gets \Gamma  \cup \{D'\}$}
		\ENDFOR

		\STATE {Train model $M$ with the whole training set $\Gamma$}
		\FOR {dataset $x$ in $\Gamma$}
		\STATE {Calculate results on the model $M$, $M(x)$}
		\STATE {$\Omega \gets \Omega \cup \{M(x)\}$}
		\ENDFOR
		\RETURN {$\Omega$}
	\end{algorithmic} 
\end{algorithm}

As the augmented data are more difficult to learn, it is natural to let the model learn the original data first and then the augmented data from easy to hard. We detail our curriculum learning strategy based on this hypothesis in the next section.

The second hypothesis we propose is to solve the multiclass classification task. Image classification, text classification like news, and algorithm classification are all classical multiclass classification tasks. The task is quite difficult, and a common simplification is to split the multiclass classification task into easily solvable subtasks with fewer classes. Hence, we formulate the hypothesis that \textbf{for the multiclass classification task, it is more effective to determine the learning order of the model from a class perspective}. Based on this hypothesis, the optimization goal of the model gradually transitions from a classification problem with few classes to a classification of multiple classes during the entire training process. Intuitively, the task is much easier to solve under this setting compared to a straightforward solution. We next conduct an experiment to verify the hypothesis.

The difficulty of code data may be reflected in the length of the code, the use of rare tokens, the complexity of logic, etc. Although these heuristics are reasonable for people, they are not necessarily the case for models. Therefore, unlike the previous validation experiment that uses code augmentation techniques to distinguish the difficulty of the samples artificially, we let the model itself give an evaluation of the data as the difficulty scores, as shown in Algorithm \ref{algorithm2}.

The purpose of Algorithm \ref{algorithm2} is to get the average difficulty score of each class on the training set. To get the difficulty score of each sample on the training set, we apply the leave-one-out strategy, \ie, when we compute the difficulty scores for a part of the samples, we train the model with all the other data. (line 4-9) Then we compute the average difficulty scores on each class. (line 11-14) 

\begin{figure}[t]
	\centerline{\includegraphics[width=\linewidth]{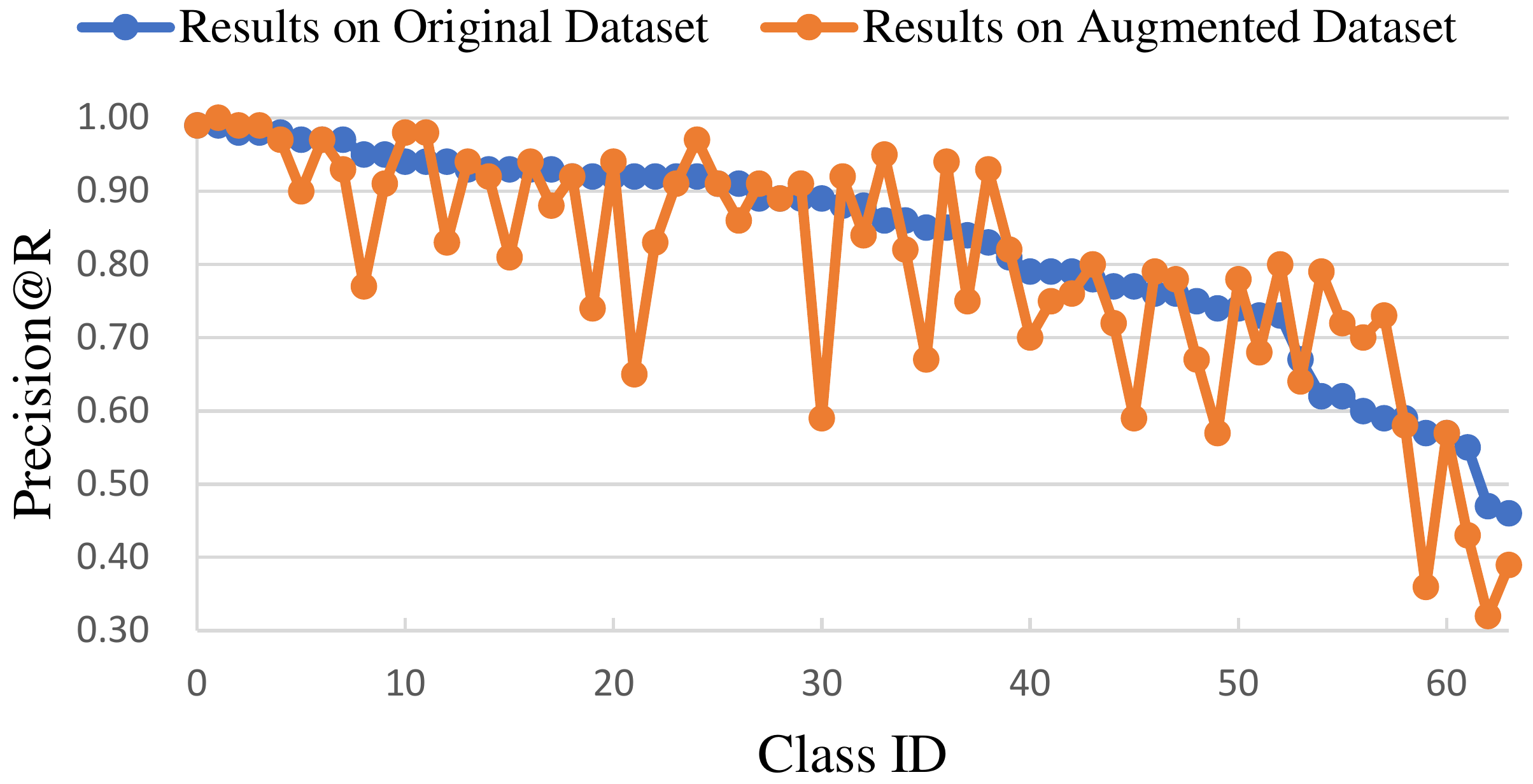}}
	\caption{Visualization of average performance across training classes for POJ104 dataset by CodeBERT.}
	\label{class_score}
\end{figure}

To have a comparison with the learning order under the first hypothesis, we also apply Algorithm~\ref{algorithm2} to the state-of-the-art model CodeBERT with POJ104 dataset. POJ104 dataset contains many classes, and the task of POJ104 dataset is to predict the class for a given program. We apply Algorithm \ref{algorithm2} to both the original training set and the augmented training set. We sort their average difficulty scores of each class according to the scores on the original training set, as shown in Fig. \ref{class_score}.

\begin{algorithm}[H]
	\caption{Validation Algorithm for Hypothesis 2}
	\label{algorithm2}
	\begin{algorithmic}[1] 
		\REQUIRE Training set $D$, the entire dataset $S$, model $M$
		
		\STATE {$C \gets \{\}$, a set of difficulty scores}
		\STATE {$\Theta \gets \{\}$, a set of average difficulty scores on classes}
		\STATE {Split training set $D$ uniformly as \{$D_i : i=1~...~N$\} }
		
		\FOR {$i \gets 1~...~N$}
		\STATE {Calculate the difference set of $D_i$ over $S$, $S-D_i$}
		\STATE {Train model $M$ with $S-D_i$ and get model $M_i$}
		\STATE {Evaluate $D_i$ with $M_i$ and obtain the experimental results of each sample as the difficulty score set $C_i$ }
		\STATE {$C \gets C  \cup C_i$}
		\ENDFOR
		
		\STATE {Train model $M$ with training set $D$}
		\FOR {class $x$ in $D$}
		\STATE {Calculate average difficulty scores on class $x$, $\mu(C,x)$}
		\STATE {$\Theta \gets \Theta \cup \{\mu(C,x)\}$}
		\ENDFOR
		\RETURN {$\Theta$}
	\end{algorithmic} 
\end{algorithm}

From Fig. \ref{class_score} it can be found that the performance of the model on various classes varies greatly. The experimental performance reflects the difficulty of classes; the better the experimental performance, the lower the difficulty, and vice versa. Also, we find that the performance on the augmented dataset is almost always lower than that on the original dataset, further validating our previous hypothesis. At the same time, Fig. \ref{class_score} shows that the performance of the model on the augmented dataset, although decreasing, is always distributed around the performance of the same class on the original dataset. Therefore, we conclude that for multiclass classification tasks organizing the data by class can yield data with more stable gradients than artificially differentiating the data by augmentation techniques. It motivates us to expose models to the easier classes first and then gradually transition to the harder classes.

\section{Proposed Approach}\label{sec3}
In this section, we describe the details of our approach. Our method is built on the fine-tuning paradigm and adapts pre-trained models to  downstream tasks. Given pre-trained models and datasets of downstream tasks, we exploit the potential of pre-trained models on these tasks by acting on the data only.
\subsection{Approach Overview}
\begin{figure}[t]                         
	\centering                              
	\includegraphics[width=\linewidth]{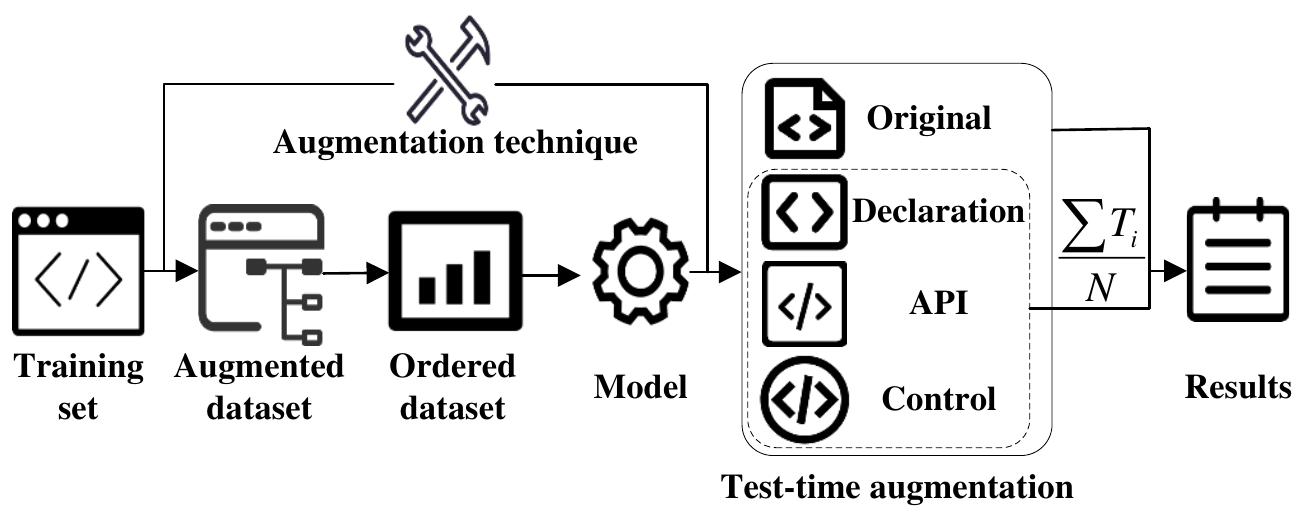} 
	\caption{Overview of our proposed method.}
	\label{pic1}                              
\end{figure}

Fig.~\ref{pic1} presents an overview of our approach. Our approach mainly consists of three components.
\begin{itemize}
	\item \textbf{Augmentation for training data} that transforms given programs into semantically equivalent programs and build augmented dataset to make training data more diverse.
	\item \textbf{Curriculum strategy} that organizes augmented dataset into the ordered dataset in an easy-to-hard order. The order is determined by scoring functions.
	\item \textbf{Test-time augmentation} that yields transformed versions of programs for prediction. The results are the fusion of results of original programs and transformed programs of different transformation types.
\end{itemize}

\begin{table}[htbp]
	\caption{Code Transformation Techniques}
	\begin{center}
		\scalebox{0.85}{
		\begin{tabular}{c|c|c}
			\hline
			\textbf{\tabincell{c}{Transformation\\ Family}}&\textbf{C/C++}&\textbf{Java}\\
			\hline
			\textbf{Control} & \tabincell{c}{for/while/if\\
				transformer}&\tabincell{c}{for/while/if\_else\\ transformer} \\ 
			\hline
			\textbf{API} &\tabincell{c}{ input/output \\
				c/cpp\_style \\transformer}
			& \tabincell{c}{equal\_loc/equal\_func/ \\
				add\_assign transformer} \\
			\hline
			\textbf{\tabincell{c}{Declaration\\and other}} & \tabincell{c} {unused\_decl/brace/ \\
				return transformer} &\tabincell{c}{stmt\_sort/merge/divide\\ transformer } \\
			\hline
			
		\end{tabular}}
		\label{tabtransformation}
	\end{center}
\end{table}

\subsection{Augmentation for Training Data}\label{AA}
In order to help models learn code features in a way that are invariant to semantically equivalent programs, we construct semantic-preserving transformations for code data. The lexical appearances and syntactical structures are different before and after transformations, but the semantics of programs are identical.

Various languages apply different transformation techniques due to specific language characteristics. In this paper, we use the same transformation techniques for data in the same language  which do not rely on prior knowledge from tasks or datasets. There are two programming languages in our experiments. For C/C++, we modify the work from Quiring \etal~\cite{Quiring2019MisleadingAA}. For Java, we apply the SPAT tool~\cite{SPAT}. We apply ten transformations for C/C++ and nine transformations for Java. The specific transformations are shown in Table~\ref{tabtransformation}. These techniques are grouped by the granularity of their changes. They change the control structure, API and declaration, respectively, to help models extract and learn the corresponding features, while ensuring that the semantics remain unchanged. Taking the transformations in Fig.~\ref{example1} as an example, the $for$ transformer is applied to transform the original program to the $1-Transformation$ program and converts the $for$ structure to $while$. This type of transformation  enables the model to understand various control structures. From $1-Transformation$ program to $2-Transformation$ program, $unused\_decl$ and $brace$ transformer are applied. This type of transformations could also generate diverse and equivalent declaration statements by merging, splitting and swapping declaration statements, helping the model to ignore the interference of syntactic formals and focus on semantics. In the last transformation to $3-transformation$ program, the output API $cout$ is converted to $printf$. The API transformation exploits the fact that the same function can be implemented by different APIs. These transformation techniques would also work in combination to make the dataset more diverse.

\subsection{Curriculum Strategy}
The key challenge of curriculum learning is how to define easy/difficult examples. In this paper, we propose two difficulty scoring functions based on the hypotheses presented in Section~\ref{hypotheses}.

\paragraph{Augmentation-based Curriculum Strategy} 

The previous section has introduced data augmentation techniques for code data, and it is cheap to generate diverse data through transformations. However, compared with original data, the augmented data can be regarded as perturbations or adversarial examples of original data~\cite{Ramakrishnan2020SemanticRO,Yefet2020AdversarialEF}, and they should be more difficult to learn as verified in Section \ref{hypotheses}. 

Therefore, we design an augmentation-based curriculum strategy. We first train on only the original data, and then gradually increase the proportion of the augmented data, ensuring that the model is exposed to more data and the difficulty gradually increases during the training process. 

In particular, it should be noted that in the process of learning the augmented data we do not strictly follow the order of $1-Transformation$ programs to $multi-Transformation$ programs, since we find that some programs have far more transformed program variants than others and multiple transformations could cause the data to be unbalanced. Therefore, we sample an equal number of augmented samples from the transformed program variants of each sample in the original training set for learning, and the data statistics are
shown in Table \ref{dataset}. This method is easy to implement on general models, and we illustrate its effects in the following experiments.

\paragraph{Class-based Curriculum Strategy}\label{classbased}
Especially for multiclass classification tasks, based on the hypothesis verified in Section \ref{hypotheses}, we propose a class-based curriculum strategy.

Specifically, the leave-one-out strategy is employed to obtain the difficulty scores on the entire training set, and then the average difficulty score on each class is calculated. The samples in the same class take the average class difficulty score as their difficulty scores. In the training process, this setting allows the model to learn easier classes first, and then to more difficult classes. Obviously, the model needs to extract and learn more features to deal with increasingly difficult tasks.

Once the scoring function is determined, we still need to define the pace at which we transition from easy samples to harder samples. With reference to the work~\cite{Penha2020CurriculumLS}, when selecting and applying different pacing functions, we ensure that the model has a number of samples to learn when the training iteration begins, and gradually gets in touch with difficult samples until all samples are available. We implement a range of pacing functions according to Penha~\etal~\cite{Penha2020CurriculumLS} and illustrate its effects in Section \ref{pacing_func}.

\subsection{Test-Time Augmentation}
We also apply augmentations to the test set. These are the same as the augmentation techniques applied on the training set. TTA neither modifies trained models nor changes test distribution. It performs predictions on the same test data as the general test procedure, except that the prediction for each test input is the aggregation of predictions on multiple transformed versions.

To further eliminate overconfident incorrect predictions due to overfitting~\cite{Wang2019AleatoricUE}, for each sample in the test set we sample three augmented copies from its transformed candidates.
Sampling more samples for prediction may make the results more robust, but would increase the prediction time proportionally. As shown in the right part of Fig.~\ref{pic1}, the final experimental performance is the sum of results on the original test set and results in the augmented perspective, which are the average of the results on augmented copies. As a result, incorrect prediction on a single test case by the model is corrected by combining multiple perspectives to make a final prediction.

\section{Experiments}\label{sec4}
In this section, we conduct experiments to verify whether our method is effective in different tasks, including algorithm classification, code clone detection and code search tasks.  

\begin{table}[htbp]
	\caption{Data Statistics}
	\begin{center}
		\scalebox{0.85}{
		\begin{tabular}{lcc}
			\hline
			\textbf{Dataset}&\textbf{Original training set}&\textbf{Augmented training set}\\
			\hline
			POJ104 & 30815 & 123058 \\ 
			CodeCloneBench & 901028 & 3362570 \\
			
			CodeSearchNet & 164923 & 331533 \\
			\hline
			
		\end{tabular}}
		\label{dataset}
	\end{center}
\end{table}

\subsection{Data preparation}\label{datasett}
In this subsection, we present benchmark datasets for three tasks from CodeXGLUE~\cite{Lu2021CodeXGLUEAM}: POJ104, BigCloneBench~\cite{Svajlenko2014TowardsAB} and CodeSearchNet~\cite{Husain2019CodeSearchNetCE} and describe how to simply adapt data of various tasks to our approach. 

POJ104 dataset is collected from an online judge platform, which consists of 104 program classes and includes 500 student-written C/C++ programs for each class. The task for POJ-104 dataset is to retrieve other programs that solve the same problem as a given program. We split the dataset according to labels. We use 64 classes of programs for training, 24 classes of programs for testing, and 16 classes of programs for validation. For data augmentation, to successfully compile the programs, ``\#include'' statements are prepended before the programs. This process does not introduce differences since added statements are the same for all programs. As some programs cannot be compiled, we further use regular expressions to correct programs with simple grammatical errors, and remove the rest with serious grammatical and semantic problems. A total of 1710 programs were removed, accounting for about 3\%~(1710/52000). To guarantee the fairness of the experiments, we also evaluate the baseline models on both the original dataset and the normalized dataset. For test-time augmentation, the results of the original and augmented versions of the same program are merged to make a prediction.

BigCloneBench dataset contains 25,000 Java projects, cover 10 functionalities and including 6,000,000 true clone pairs and 260,000 false clone pairs. The dataset provided by Wang~\etal~\cite{Wang2020DetectingCC} is filtered by discarding code fragments without any tagged true or false clone pairs, leaving it with 9,134 Java code fragments. The dataset includes 901,028/415,416/415,416 pairs for training, validation and testing, respectively. This dataset has been widely used for the code clone detection task. For code augmentation, since the data is in the form of code pairs, we replace any original program in clone pairs with augmented programs to form new pairs. For test-time augmentation, all versions of a code pair are considered to determine whether it is a clone pair.

CodeSearchNet contains about 6 million functions from open-source code spanning six programming languages. In this paper, we use the dataset in Java. Given a natural language query as the input, the task is to find the most semantically related code from a collection of candidate programs. According to the state-of-the-art model GraphCodeBERT~\cite{Guo2021GraphCodeBERTPC}, we expand 1000 query candidates to the whole code corpus, which is closer to the real-life scenario. The answer of each query is retrieved from the whole validation and testing code corpus instead of 1,000 candidate programs. For code augmentation in the training set, since the data are pairs of natural language queries and programming language fragments, we replace original programs with augmented programs and form new pairs with their natural language queries. When doing test-time augmentation, it is different from the previous two tasks. Since the test set is the set of natural language queries, we apply code augmentation techniques to the codebase corresponding to these queries. For a query and each code in codebase, we calculate similarity of the code and its multiple transformed versions to the query, respectively. We use the average of similarity for sorting and evaluation.

The original and augmented data statistics of the above tasks are shown in Table ~\ref{dataset} and the augmented datasets contain the original data. We release all data for verification and future development. Theoretically, more augmented data can be obtained, however, more data to train would bring larger time overhead. To trade off the experimental performance and time overhead, we use a limited amount of augmented data, and we apply curriculum learning strategy where the model is trained from a smaller data size and the overhead is further reduced.

\subsection{Experimental Setups}

To illustrate the effectiveness of our method on code-related tasks, we build our approach on code pre-trained models CodeBERT and GraphCodeBERT. To illustrate the applicability of our method, we also evaluate our method on natural language pre-trained model RoBERTa~\cite{Liu2019RoBERTaAR} that has not been exposed to code at all. In replication experiments, we follow the description in their original papers and released code. For parameter settings, to ensure fairness, we keep all parameters consistent with their released code including random seeds except for the warmup step and epoch. The warmup step parameter adapts to the increase of the dataset, and its value is adjusted from the original dataset size to the augmented dataset size. Also due to the increase in data size and the progressive curriculum learning, we increase the epoch and set it to 20, 10, and 15 on POJ104, BigCloneBench, and CodeSearchNet, respectively. We replicate CodeBERT and GraphCodeBERT with the same parameter settings. The results reported in the original papers and our replicated results are not much different, and we present all the results. For data augmentation, we implement augmentaion techniques on the top of Clang~\cite{clangurl} for C/C++. With respect to pacing function, the hyperparameters are set according to Penha~\etal~\cite{Penha2020CurriculumLS}. 

\subsection{Algorithm Classification}
\paragraph{Metrics and Baselines} 
We use precision and MAP as the evaluation metrics of the algorithm classification task. Precision is defined as the average precision score and MAP is the rank-based mean of average precision score, each of which is evaluated for retrieving most similar samples given a query. We apply RoBERTa and the state-of-the-art model CodeBERT as baseline methods. RoBERTa is a pre-trained model on natural language. CodeBERT is a pre-trained model on code data. It combines masked language modeling~\cite{Devlin2019BERTPO} with replaced token detection objective~\cite{Clark2020ELECTRAPT} to pre-train a Transformer~\cite{Vaswani2017AttentionIA} encoder.

\begin{table}[htbp]
	\caption{Algorithm Classification Comparison}
	\begin{center}
		\begin{tabular}{lcc}
			\hline
			\textbf{Model}&\textbf{Precision}&\textbf{MAP}\\
			\hline
			RoBERTa & 82.82 & 80.31(76.67) \\ 
			RoBERTa + DA + CL & 88.15 & 86.55 \\
			\hline
			CodeBERT & 85.28 & 82.76(82.67) \\
			CodeBERT + DA + CL & \textbf{93.63} & \textbf{92.91} \\
			\hline
			
		\end{tabular}
		\label{algorithm_result}
	\end{center}
\end{table}

\paragraph{Results} 
We compare with and without our method (DA + CL) for these pre-trained models. Table~\ref{algorithm_result} summarizes these results. For baseline methods, all experimental results are evaluated on our normalized dataset, except for results of MAP in parentheses. These results are reported in the original paper of baseline methods and MAP is their only metric for algorithm classification task. Natural language pre-trained model RoBERTa fine-tuned with our method, achieves 88.15\% on precision, 86.55\% on MAP. Our method improves its performance noticeably by 5.33\% on precision, 6.31\% on MAP and 9.88\% compared to the results reported in the original paper. Code pre-trained model CodeBERT fine-tuned with our method, achieves 93.63\% precision and 92.91\% on MAP. Our method substantially improves 8.35\% on precision, 10.15\% on MAP, and 10.24\% compared to the original result. Notably, with our method, RoBERTa model without being pre-trained on code data outperforms the existing state-of-the-art model CodeBERT fine-tuned on this task by 3.79\% MAP.
\subsection{Code Clone Detection}
\paragraph{Metrics and Baselines} 
We use precision, recall and F1 score as the evaluation metrics of the code clone detection task. In our experiments, we compare a range of models including the state-of-the-art model GraphCodeBERT. GraphCodeBERT is a pre-trained model for code which improves CodeBERT by modeling the data flow edges between code tokens. CDLH~\cite{Wei2017SupervisedDF} learns representations of code fragments through AST-based LSTM. ASTNN~\cite{Zhang2019ANN} encodes AST subtrees for statements and feeds the encodings of all statement trees into an RNN to learn representation for a program. FA-AST-GMN~\cite{Wang2020DetectingCC} leverages explicit control and data flow information and uses GNNs over a flow-augmented AST to learn representation for programs. TBCCD~\cite{Yu2019NeuralDO} proposes a tree convolution-based method to detect semantic clone, that is, using AST to capture structural information and obtain lexical information from the position-aware character embedding.  

\begin{table}[htbp]
	\caption{Code Clone Detection Comparison}
	\begin{center}
		\scalebox{0.85}{
		\begin{tabular}{lccc}
			\hline
			\textbf{Model}&\textbf{Precision}&\textbf{Recall}&\textbf{F1}\\
			\hline
			CDLH & 0.92 & 0.74 & 0.82 \\ 
			ASTNN & 0.92 & 0.94 & 0.93 \\
			FA-AST-AMN & 0.96 & 0.94 & 0.95 \\
			TBBCD & 0.94 & 0.96 & 0.95\\
			\hline
			RoBERTa(10\% data) & 0.966 & 0.962 & 0.964(0.949)\\
			RoBERTa(10\% data) + DA + CL & 0.973 & 0.957 & 0.965\\
			\hline
			CodeBERT(10\% data) & 0.960 & 0.969 & 0.965\\
			CodeBERT(10\% data) + DA + CL & 0.972 & \textbf{0.972} & \textbf{0.972}\\
			\hline
			GraphCodeBERT & \textbf{0.973} & 0.968 & 0.971\\
			\hline
			
		\end{tabular}}
		\label{clone_result}
	\end{center}
\end{table}
\paragraph{Results} 
Table~\ref{clone_result} shows results for code clone detection. Our reproduced results are mostly consistent with results reported in original papers, except for the F1 score of 0.964 for RoBERTa, which is higher than the original result of 0.949. We implement our method on RoBERTa and CodeBERT. Experiments show that models with our method consistently perform better than the original models. Notably, with our method, RoBERTa performs comparably to CodeBERT, and CodeBERT outperforms the state-of-the-art model GraphCodeBERT. More importantly, following the original settings of CodeBERT, CodeBERT only randomly samples 10\% of the data for training compared to GraphCodeBERT. Even though we expand the data using data augmentation in the experiment for CodeBERT, the data used by CodeBERT are still much less than data for GraphCodeBERT.
\subsection{Code Search}

\paragraph{Metrics and Baselines} 
For code search task, we use MRR as the evaluation metric. MRR is the average of the reciprocal rank of results of a set of queries. The reciprocal rank of a query is the inverse of the rank of the first hit result.

Table~\ref{codesearch_result} shows the results of different approaches on the CodeSearchNet corpus.  The first four rows are reported by Husain \etal~\cite{Husain2019CodeSearchNetCE}. NBOW, CNN, BIRNN and SELFATT represent neural bag-of-words~\cite{Sheikh2016LearningWI}, 1D convolutional neural network~\cite{Kim2014ConvolutionalNN}, bidirectional GRU-based recurrent neural network~\cite{Cho2014LearningPR}, and multi-head attention~\cite{Vaswani2017AttentionIA}, respectively.

\begin{table}[htbp]
	\caption{Code Search Comparison}
	\begin{center}
		\begin{tabular}{lc}
			\hline
			\textbf{Model}&\textbf{MRR}\\
			\hline
			NBow & 0.171 \\ 
			CNN & 0.263 \\ 
			BiRNN & 0.304 \\ 
			SelfAtt & 0.404 \\ 
			\hline
			RoBERTa & 0.599 \\ 
			RoBERTa(code) & 0.620 \\ 
			RoBERTa + DA + CL & 0.635  \\
			\hline
			CodeBERT & 0.676  \\
			CodeBERT + DA + CL & 0.697 \\
			\hline
			GraphCodeBERT & 0.696(0.691)  \\
			GraphCodeBERT + DA + CL & \textbf{0.720} \\
			\hline
			
		\end{tabular}
		\label{codesearch_result}
	\end{center}
\end{table}
\paragraph{Results} 
Table~\ref{codesearch_result} shows results of different approaches for code search. RoBERTa~(code) is pre-trained on programs from CodeSearchNet with masked language modeling while maintaining the RoBERTa architecture. Our reproduced result 0.696 of GraphCodeBERT is slightly differently from the originally reported result 0.691. We implement our method on RoBERTa, CodeBERT and the state-of-the-art model GraphCodeBERT for code search. The results show that natural language pre-trained model RoBERTa with our method outperforms RoBERTa (code), which is the same model architecture pre-trained on code data. CodeBERT with our method outperforms the original state-of-the-art model GraphCodeBERT. The performance of GraphCodeBERT with our method reaches 0.720 MRR, surpassing the original result 0.691 MRR.
\subsection{Summary}
On above tasks and their benchmark datasets, our method substantially improves the performance of a range of pre-trained models, achieving the state-of-the-art performance on all tasks. For the natural language pre-trained model with no exposure to code at all, with the help of our approach, it is able to match or even surpass existing code pre-trained models normally fine-tuned to corresponding tasks. In the code search task, RoBERTa pre-trained with natural language and fine-tuned with our method, surpasses the same architecture pre-trained with code data and fine-tuned with the general method. These all illustrate the strong bridging role of our method between pre-trained models and code-related downstream tasks by introducing semantic knowledge for downstream tasks into pre-trained models.

For code-related tasks, applying our approach to a pre-trained model at the finetune stage with a relatively small cost is preferable to pre-training a more complicated model from scratch with huge resources. It illustrates the superiority of our method, but this is not to negate the work of code pre-trained models either. In fact, our approach achieves better results when applied to a superior  pre-trained model. Probably, the research of pre-trained models for source code has much work to do in terms of data diversity and conjunction with downstream tasks.

\section{Analysis}\label{sec5}
This section analyzes the effects of different parameters on the performance of tasks in our experiment.

\subsection{Ablation Study}
This section investigates how data augmentation and curriculum learning affect the performance of models, respectively. The following subsections show these results for algorithm classification, code clone detection and code search task.

\begin{table}[htbp]
	\caption{Ablation Study on Algorithm Classification}
	\begin{center}
		\begin{tabular}{lcc}
			\hline
			\textbf{Model}&\textbf{Precision}&\textbf{MAP}\\
			\hline
			CodeBERT & 85.28 & 82.76 \\ 
			\hline
			CodeBERT + DA + CL & 93.63 & 92.91 \\
			
			w/o DA-Training & 91.90 & 90.79 \\
			w/o TTA & 88.76 & 87.21 \\
			w/o CL & 92.55 & 91.52 \\
			\hline
			
		\end{tabular}
		\label{algorithm_ablation}
	\end{center}
\end{table}

\paragraph{Algorithm Classification} 
For algorithm classification task, we conduct experiments without augmention on training set (DA-Training), test-time augmentation or curriculum learning. The results are shown in Table~\ref{algorithm_ablation}. The first row shows the results of the baseline model. The second row presents the results of the baseline model with our full method. The third row removes augmentation on the training set. The fourth row presents the results of removing test-time augmentation. The results of removing curriculum learning strategy are shown in the last row. As seen from the results, removing any of the components leads to a drop of the model performance, and the removal of test-time augmentation leads to a significant performance degradation, indicating that all three components are necessary to improve performance, and test-time augmentaion contributes the most to the improvements. We believe that for clustering tasks similar to algorithm classification, integrating multiple perspectives in a data augmentation manner during testing could be a huge boost to model performance.

\begin{table}[htbp]
	\caption{Ablation Study on Code Clone Detection}
	\begin{center}
		\begin{tabular}{lccc}
			\hline
			\textbf{Model}&\textbf{Precision}&\textbf{Recall}&\textbf{F1}\\
			\hline
			CodeBERT & 0.963 & 0.965 & 0.964 \\ 
			\hline
			CodeBERT + DA + CL & 0.972 & 0.972 & 0.972 \\	
			w/o TTA & 0.971 & 0.972 & 0.971 \\
			w/o CL & 0.976 & 0.965 & 0.970 \\
			w/o DA-Training + CL & 0.964 & 0.965 & 0.964 \\
			\hline
			
		\end{tabular}
		\label{clone_ablation}
	\end{center}
\end{table}
\paragraph{Code Clone Detection} 
For code clone detection task, we also conduct experiments without augmention on training set, test-time augmentation or curriculum learning. Unlike algorithm classification, we apply augmentation-based curriculum learning for code clone detection task. The removal of augmentation on the training set means that the CL component also does not work, and only test-time augmentation component works. The experimental results in Table~\ref{clone_ablation} show that the combination of augmentation on the training set and CL component has the largest performance improvement, and test-time augmentation has no significant performance improvement, but the model can still benefit from it.

\begin{table}[t]
	\caption{Ablation Study on Code Search}
	\begin{center}
		\begin{tabular}{lc}
			\hline
			\textbf{Model}&\textbf{MRR}\\
			\hline
			GraphCodeBERT & 0.696  \\ 
			\hline
			GraphCodeBERT + DA + CL & 0.720 \\	
			w/o TTA & 0.707 \\
			w/o CL & 0.708 \\
			w/o DA-Training + CL & 0.710 \\
			\hline
			
		\end{tabular}
		\label{search_ablation}
	\end{center}
\end{table}

\paragraph{Code Search} 
With the same ablation experimental setups as for the code clone detection task, we conduct experiments on the code search task. As shown in Table~\ref{search_ablation}, we conclude that all three components are necessary for the improvements. The last row shows the result using only test-time augmentation, which is able to significantly exceed the original state-of-the-art performance without training with additional augmentation data. We speculate that test-time augmentation is able to combine multiple augmentation copies in the code retrieval process to make judgments and eliminate overconfident incorrect predictions on the original test set. The penultimate row shows the experimental result of removing CL component. In other words, it is obtained by the combination of augmentation on the training set and test-time augmentation acting on the model. Compared to the result of applying test-time augmentation component only in the last row, we find that more augmented data used for training may result in negative gains. One possible reason is that the augmented data introduces more noise, causing the model to choose from more candidates for the same query during training. These results further illustrate the necessity of curriculum learning on augmented data.

\begin{table}[htbp]
	\caption{Effects of Augmentation Types on Algorithm Classification}
	\begin{center}
		\begin{tabular}{lcc}
			\hline
			\textbf{Model}&\textbf{Precision}&\textbf{MAP}\\
			\hline
			All & 93.63 & 92.91 \\
			w/o Declaration & 92.13 & 90.88 \\
			w/o API & 92.35 & 91.24 \\
			w/o Control & 94.17 & 93.41 \\
			\hline
			
		\end{tabular}
		\label{augmentation_ablation}
	\end{center}
\end{table}

\subsection{Effects of Augmentation Type}
Since this paper considers multiple augmentation techniques, in this section we explore the effects of augmentation techniques at different granularities on the experimental results. We build transformed datasets of the same size using augmentation techniques of different granularities and train CodeBERT separately on these datasets for algorithm classification task. Results are shown in Table~\ref{augmentation_ablation}. The first row shows the results using all augmentation techniques of three granularities, while the second to fourth rows show the results without the augmentation techniques for the declaration, API, or control stucture granularity, respectively. From the results, it can be seen that not using the augmentation techniques of declaration or API granularity leads to a decrease in results, while not using the augmentation techniques of control sturcture leads to an increase. This indicates that the augmentation of declaration and API contribute more to the improvements, however, the control structure augmentation introduces more noise than contribution. We speculate that changing the control structure has a greater impact on the token order and context relative to the other two granularities of augmentation techniques, and pre-trained models we use are based on masked language modeling and are context sensitive. These reasons make it more difficult for the models to learn the knowledge and features introduced in the process of changing the control structure. This finding also encourages the code pre-trained model to further exploit structural information of source code in order to better understand the program semantics.

\begin{table}[htbp]
	\caption{Effects of Pacing Function on Algorithm Classification}
	\begin{center}
		\begin{tabular}{lcc}
			\hline
			\textbf{Model}&\textbf{Precision}&\textbf{MAP}\\
			\hline
			Random(baseline) & 85.28 & 82.76 \\
			\hline
			Anti & 81.87 & 78.73 \\
			\hline
			Linear & 86.94 & 84.96 \\
			Step & 86.39 & 84.35 \\
			Geom\_progression & 85.97 & 83.70 \\
			Root\_2 & 86.98 & 84.96 \\
			Root\_5 & 86.11 & 83.95 \\
			Root\_10 & 87.38 & 85.33 \\
			\hline
			
		\end{tabular}
		\label{pacing_ablation}
	\end{center}
\end{table}

\subsection{Effects of Pacing Function}\label{pacing_func}
To understand how the model is impacted by the pace we go from easy to hard examples, we evaluate the effects of different pacing functions on the experimental results, as shown in Table~\ref{pacing_ablation}. We conduct experiments on POJ104 dataset in the algorithm classfication task. The learning order is determined by the scoring function described in Section \ref{classbased}. The baseline model CodeBERT is trained in a random order and the Anti method orders training samples from hard to easy. The other methods learn training samples from easy to hard, with the difference that at each epoch a different proportion of the training data are fed to the model as determined by their functions. We briefly introduce different pacing functions and the details are described in Penha~\etal~\cite{Penha2020CurriculumLS}. The $Linear$ function linearly increases the percentage of training data input to the model. $Step$ function divides training data into several groups, and after fixed epoches a group of training samples will be added for model training. $Root\_n$ and $Geom\_progression$ functions correspond to two extreme cases. $Root\_n$ function feeds the model with a large number of easy samples and then slowly increases the proportion of hard samples, while $Geom\_progression$ function does the opposite. In the $Root\_n$ function, $n$ is the hyperparameter, and the larger the value of $n$, the more training data are fed to the model at the beginning. All these functions are fed with the same training data at the final stage of training.

In Table~\ref{pacing_ablation}, we can see that feeding data from easy to hard has a certain performance improvement, while the performance of inputting training samples from hard to easy is significantly worse than the baseline in a random order. These results illustrates the effectiveness of our curriculum learning strategy and scoring functions. Comparison of different pacing functions shows that $Linear$ and $Step$ functions achieve similar results as $Root\_2$ function. The $Root$ functions obviously outperform the $Geom\_progression$ function, which is consistent with the findings of Sohrmann ~\etal~\cite{Sohrmann2020NationwideIO} and Penha~\etal~\cite{Penha2020CurriculumLS}. The reasons are that the root function gives the model more time to learn from harder instances and is better than no CL in terms of statistical significance. In our experiments, we used $Root\_10$ function for algorithm classification task, and since we did not perform ablation study on the datasets of the other two tasks, we use $Linear$ function by default. The performance on these two tasks could probably be further improved with different pacing functions, and we leave it for future work.

\section{Related Work}\label{sec6}

\subsection{Data Augmentation}
Data augmentation aims to increase the data diversity and thus the generalization ability of the model by various transformation techniques. This approach is widely used in the computer vision domain~\cite{Wei2019GenerativeIT,Shorten2019ASO,Zhong2020RandomED}. In recent years, researchers apply data augmentation to code data as well~\cite{Quiring2019MisleadingAA,Ramakrishnan2020SemanticRO,Zhang2020GeneratingAE,Yefet2020AdversarialEF,Rabin2020OnTG}. A series of studies are motivated by the fact that existing models are vulnerable to adversarial examples, and they design methods to expose the vulnerability of models and improve the robustness of models. Our aim is to make the models more generalizable and perform better on real data, unlike the methods described above. Jain~\etal~\cite{Jain2020ContrastiveCR} improve accuracy in code summarization and type inference task based on equivalent data transformations and unsupervised auxiliary tasks. Nghi~\etal~\cite{Bui2021SelfSupervisedCL} propose a self-supervised contrastive learning framework for code retrieval and code summarization tasks. Our aim is similar to these studies, but we do not need to design the objective function or model architecture. Without the need for complicated model design, our approach accomplishes the same goal by acting on the data only. We simply augments the data and feeds the augmented data into the model in an easy-to-hard manner. Therefore, our lightweight method can be easily applied over existing models and various downstream tasks.

\subsection{Curriculum Learning}
Learning educational material in order from easy to difficult is very common in the human learning process. Inspired by cognitive science~\cite{Rohde1999LanguageAI}, researchers have found that model training can also benefit from a similar curriculum learning setting. Since then, CL has been successfully applied to image classification~\cite{Gong2016MultiModalCL,Hacohen2019OnTP}, machine translation~\cite{Kocmi2017CurriculumLA,Platanios2019CompetencebasedCL,Zhang2018AnEE}, answer generation~\cite{Liu2018CurriculumLF} and information retrieve~\cite{Penha2020CurriculumLS}.

The core of CL lies in the design of the scoring function, that is, how to define easy and hard samples. A straightforward approach is to study the data to create specific heuristic rules. For example, Bengio~\etal~\cite{Bengio2009CurriculumL} use images containing less varied shapes as easy examples to be learned first. Tay~\etal~\cite{Tay2019SimpleAE} use paragraph length as an evaluation criterion for difficulty in the question answer task. However, these are highly dependent on the task dataset and cannot be generalized to general tasks. Guo~\etal~\cite{Guo2018CurriculumNetWS} examine the examples in their feature space, and define difficulty by the distribution density, which successfully distinguishes noisy images. Xu \etal~\cite{Xu2020CurriculumLF} generally distinguish easy examples from difficult ones on natural language understanding tasks by reviewing the training set in a crossed way. In this paper, similar to Xu \etal~\cite{Xu2020CurriculumLF}, we also utilize cross validation to measure data difficulty by model itself, but we also take the class distribution into consideration. We intuitively solve the multiclass classification problem from a class perspective by first transforming it into a classification of fewer easy classes and then gradually increasing the number of difficult classes. At the same time, we combine curriculum learning and data augmentation to overcome the problem that augmented data is more difficult to learn. We first learn the original data, then gradually transition to augmented data, and experimentally illustrate and verify the effectiveness of the design.

\section{Threats to Validity}\label{sec7}
There are several threats to validity of our method.
\begin{itemize}
	\item Due to the use of test-time augmentation in our method, this component cannot be easily applied to code generation tasks. Augmentation on the training set and curriculum learning are still applicable, e.g., Jain~\etal~\cite{Jain2020ContrastiveCR} have achieved good performance on the code summarization task using code augmentation.
	\item The transformation techniques we use are not representative of the whole. Due to the characteristics of various tasks and datasets, some transformations may lead to large improvements and some may bing no improvements. Therefore, we release the datasets for replication and reducing experimental bias. Our approach is designed to be a lightweight component that generalizes to multiple downstream tasks. For specific downstream tasks, new augmentation techniques can also be applied to optimize the performance.
	\item  Due to limited computed resource, we did not explore the performance of our approach for the code clone detection task on GraphCodeBERT or conduct ablation stuies on all three tasks regarding the pacing function and transformation type. In fact, there should be room for improvement and interesting conclusions to be explored. We shall get better results by searching for more suitable pacing functions and transformation types for the other two tasks. We leave it for future works.
	
\end{itemize}
\section{Conclusion}\label{sec8}

In this paper, we focus on bridging pre-trained models and code-related downstream tasks and propose a lightweight approach on the fine-tuning paradigm, which is easy to implement on top of various models. We build our approach on code pre-trained models of CodeBERT and GraphCodeBERT, and these models substantially outperform original models and achieve the state-of-the-art performance on algorithm classification, code clone detection and code search. Moreover, we apply our method to natural language pre-trained model RoBERTa and it achieves comparable or better performance than existing state-of-the-art code pre-trained models fine-tuned on these tasks. This finding reveals that there is still much room for improvement in existing pre-trained models for source code understanding.

This paper focuses on code discriminative tasks. It is more challenging to apply our approach to code generation tasks. However, generation tasks are data-hungry and may require more diverse data for learning, such as code generation where multiple code candidates are expected to be generated. In the future, it would be interesting to combine our approach and prompt-based learning~\cite{Liu2021PretrainPA} to further exploit the potential of generative pre-trained models on code generation tasks.

\begin{acks}
	The authors would like to thank the anonymous reviewers for their insightful comments. This work was substantially supported by National Natural Science Foundation of China (No. 61690203, 61872373, 62032019, and U1936213). This work was also supported by the Major Key Project of PCL.
\end{acks}

\bibliographystyle{ACM-Reference-Format}
\bibliography{sample-base}


\end{document}